\documentclass[aip,amsmath,amssymb,reprint]{revtex4-1}

\usepackage{graphicx}
\usepackage{dcolumn}
\usepackage{bm}
\usepackage[utf8]{inputenc} \usepackage[T1]{fontenc} \usepackage{mathptmx} \usepackage{etoolbox}
\usepackage{array}

\usepackage{nccmath}
\usepackage{hyperref}
\usepackage{enumitem}

\makeatletter \def\@email#1#2{%
  \endgroup \patchcmd{\titleblock@produce} {\frontmatter@RRAPformat}
  {\frontmatter@RRAPformat{\produce@RRAP{*#1\href{mailto:#2}{#2}}}\frontmatter@RRAPformat} {}{} }%
\makeatother

\begin{document}
\title{Spin and velocity correlations in a confined two-dimensional fluid of disk-shaped active
  rotors}

    \author{Miguel \'Angel L\'opez-Casta\~no}
    \affiliation{Departamento de F\'isica, Universidad de Extremadura, Avda. Elvas s/n, 06071 Badajoz, Spain}
    
    \author{Alejandro M\'arquez Seco}
    \affiliation{Departamento de F\'isica, Universidad de Extremadura, Avda. Elvas s/n, 06071 Badajoz, Spain}
    
    \author{Alicia M\'arquez Seco}
    \affiliation{Departamento de F\'isica, Universidad de Extremadura, Avda. Elvas s/n, 06071 Badajoz, Spain}
    
    \author{Álvaro Rodríguez-Rivas}
    \affiliation{Department of Physical, Chemical and Natural Systems, Pablo de Olavide University, 41013, Sevilla, Spain}
    
    \author{Francisco Vega Reyes}
    \email{fvega@eaphysics.xyz}
    \affiliation{Departamento de F\'isica and Instituto de
      Computaci\'on Cient\'ifica Avanzada (ICCAEx), Universidad de Extremadura, Avda. Elvas s/n,
      06071 Badajoz, Spain}

\date{\today}

\begin{abstract}
  We study the velocity autocorrelations in an experimental configuration of confined
  two-dimensional active rotors (disks). We report persistent small scale oscillations in both
  rotational and translational velocity autocorrelations, with their characteristic frequency
  increasing as rotational activity increases.  While these small oscillations are qualitatively
  similar in all experiments, we found that, at strong particle rotational activity, the large scale
  particle spin fluctuations tend to vanish, with the small oscillations around zero persisting in
  this case, and spins remain predominantly and strongly anti-correlated at longer times. For weaker
  rotational activity, however, spin fluctuations become increasingly larger and angular velocities
  remain de-correlated at longer times. We discuss in detail how the autocorrelation oscillations
  are related to the rotational activity and why this feature is arguably should generically signal
  the emergence of chirality in the dynamics of a particulate system.
\end{abstract}

\maketitle

\section{Introduction} 
\label{intro}

Active matter comprises units that convert either internal or external energy into motion with a
systematically directed component
\cite{Dombrowski2004,Marchetti2013,Zhang2010,Petroff2015,XHZW22}. This directional, active, motion
induces a parity and time inversion symmetry break. Because of this, active matter frequently
exhibits an intrinsic and rich collective phenomenology, not present in systems of pasive particles
\cite{Bowick_2022}. For instance, the dynamics of these systems displays spontaneous
self-organization \cite{Nguyen2014}, peculiar phase separation (induced by active motion, and for
this reason known as \textit{motility induced phase separation}, or MIPS)
\cite{Cates2015,Caporusso2020,Grosmann2020}, flocking and swarming
\cite{Liao2020,GASR20,Zhang2010,Bricard2015}, and flow chirality (flows that develop patterns with
chiral asymmetry) \cite{Zhang2020,LMMRV22,Mokhtari2017,Farhadi2018,Workamp2018}. In the case of
chiral flows, its origin lies most frequently in the geometrical chirality of active particles that
compose the system \cite{Tsai2005}.


A number of fundamental aspects of the dynamics of active rotors presents very distinctive features
when compared to other equilibrium and non-equilibrium fluids. For instance, transport phenomena
display in chiral fluids a much more complex behavior, due to the emergence of antisymmetric
components in the fluxes (of mass, momentum and energy) \cite{Banerjee2017,HEM21,Han2021}. As a
consequence, the constitutive relations for these antisymmetric components of the fluxes involve the
definition of new set of transport coefficients with special properties. As an example, fluid flow
and diffusion are controlled by the so-called odd viscosity \cite{A98,Banerjee2017,Han2020} and odd
diffusion \cite{HEM21,VLR22} coefficients, both of these coefficients being specific of chiral
fluids. In the case of a fluid of active rotors, the chirality of the fluid is due to particle
systematic rotation (which, on the other hand, may have different origins). Thus, it is in this
systematic rotation where we may identify the origin of the peculiar transport properties of fluids
of active rotors. Due to this, the statistical properties of particle rotations can be identified as
an essential feature for this kind of chiral fluids and thus, one can reasonably expect the angular
velocity statistical correlations to be one of the essential factors in the dynamics of the chiral
fluid.

However, there is still little to no bibliography (to our knowledge) on the study of statistical
correlations in active rotors, specially with regard to angular velocities (for two very recent
studies on translational velocity correlations, in chiral fluids, see \cite{Zhang2020,DLJ23}). In
summary, the understanding of these correlations in chiral flows is still very preliminary. In fact,
results in a previous work suggest \cite{LMMRV22} that the structure of statistical correlations is
an essential factor determining the fluid flow properties in a system of active rotors.

Thus, the motivation of the present work is to perform a detailed and specific analysis of the
statistical correlations of angular velocities in a two-dimensional fluid of active rotors. For this
objective, we study an experimental system composed of macroscopic and flat disk-shaped
rotors. These disks are provided with 14 equal size tilted blades (see Figure~\ref{fig:processing}
a). An air upflow passes past the disks, which induces a Brownian-like translational stochastic
motion \cite{Ojha2004} and, at the same time, creates a biased particle rotation due to the presence
of the tilted blades.

For our purpose, we set-up an air table that aids fluidization in a system of 3D printed disks. An
ad-hoc fan provides stable continuous current, that is on purpose homogenized before impinging the
particles from below. As a consequence, particles continuously undergo a systematic torque $\tau_0$
which inforces them to set back a particular value of average spin after momentum-transfer events
(which in this case are either particle-particle or particle-boundary collisions). This yields a
systematic rotational activity to the rotors, as we will explain. As a consequence, the system
features peculiar statistical properties of the translational and angular velocities. In particular,
we show there are two regimes of the angular velocity correlations (at strong and weak fluidization)
for which spin autocorrelations are stronger and weaker respectively. However, translational
velocities tend to decorrelate at intermediate fluidization whereas autocorrelations are strong at
both weak and strong fluidization. We also report the existence of persistent oscillations in the
autocorrelation function of particle spin.  We describe in more detail in the remainder of this
work the features of statistical velocity correlations in our system.

This work is organized as follows. In section~\ref{sec:setup} we provide details on the experimental
configuration used for this work, along with a description of the algorithm we wrote for particle
angle tracking which, as we will explain, achieves a very high time accuracy. Additional details
related to this section, including a detailed analysis of experimental errors, can be found in the
Appendix. In Section \ref{sec:results} we show the data obtained from experiments, and analyze the
results focusing on the translational and angular velocity correlations. Finally, in
Section~\ref{sec:conclusions} , we discuss and summarize the results.

\section{Experimental set-up}
\label{sec:setup}

\subsection{Description}
\label{description}

\begin{figure}[ht]
    \begin{tabular}{m{0.35\columnwidth} m{0.65\columnwidth} }
      \includegraphics[width=0.3\columnwidth]{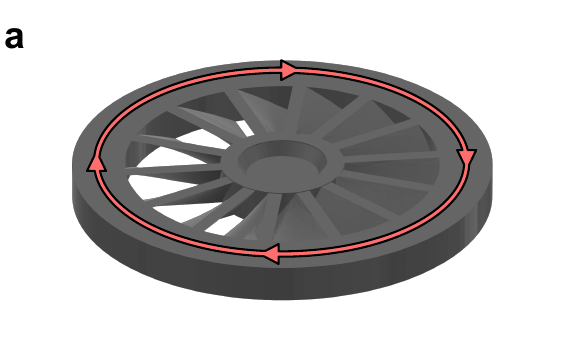} \vspace*{1.5cm}& \includegraphics[width=0.525\columnwidth]{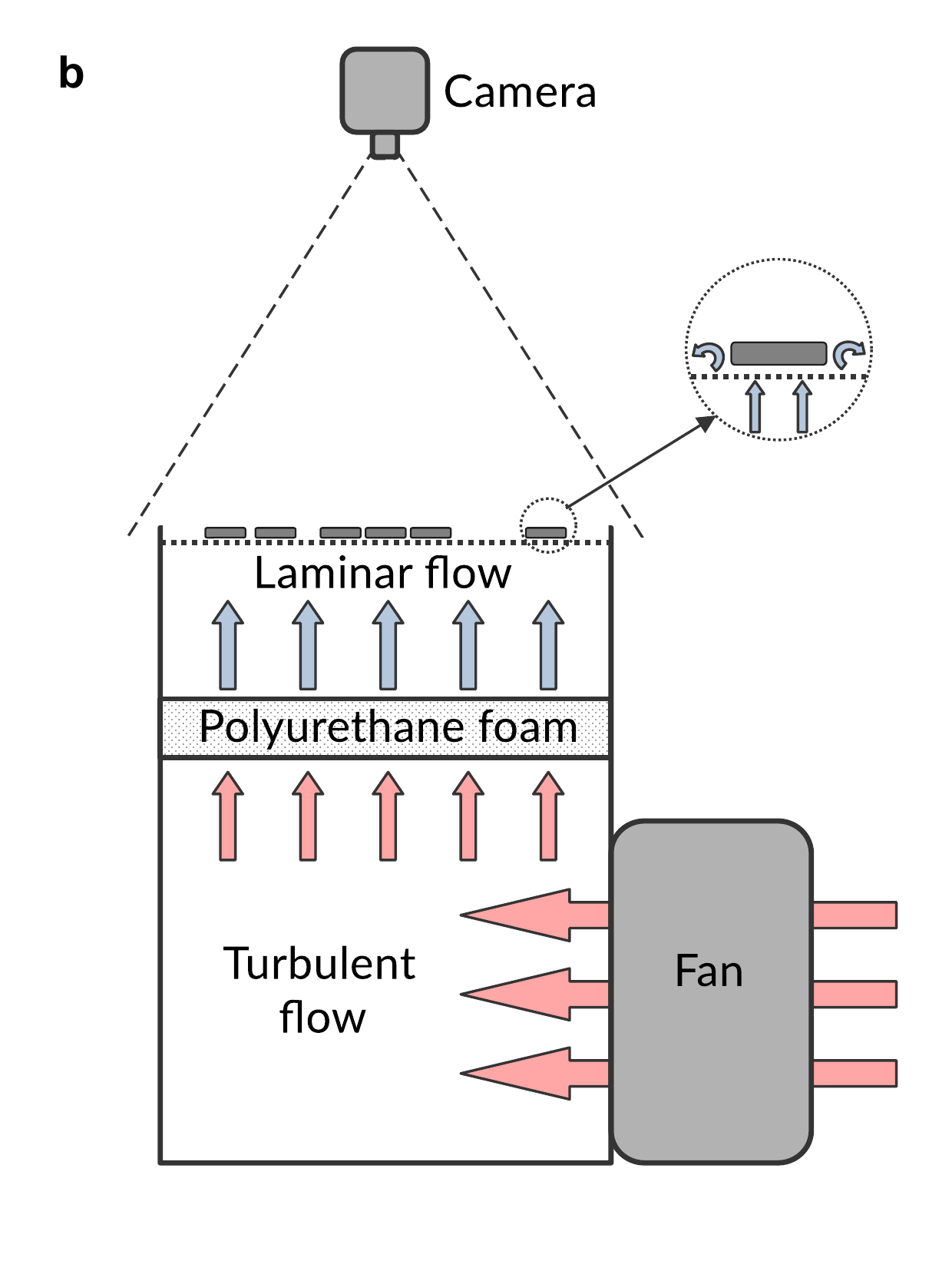} \\
    \end{tabular}
    \caption{(a) Renderization of the 3D printed disk, with its 14 blades. (b) Representation of the
      experimental set-up.}
    \label{sketch}
  \end{figure}

The fluid consists of a set of $N$ identical disk-shaped particles. Their diameter is
$\sigma=72.5 \pm 0.1~\mathrm{mm}$ and their thickness is $h = 6.0 \pm 0.1~\mathrm{mm}$. Thus, they
can be considered nearly flat (i.e., two-dimensional). Their mass is $m_p=7.76 \pm 0.01~\mathrm{g}$,
and they were produced by means of additive 3D printing (material is polylactic acid). The disks
incorporate a set of $N_b=14$ equally-spaced identical blades (reproducible 3D models, in .stl
format, are available upon request). We measured the mass and size distribution of our set of disks,
in order to control the reproducibility of the printing process. The moment of inertia of the
particles, $I$ is considered to be approximately that of a homogeneous disk.  The particle design is
rendered in Fig.~\ref{sketch}(a).


The disks are placed over a perforated square metallic grid ($100\times100~\mathrm{cm}$) and are
enclosed inside a circular wall. This wall is centered in the square grid, with height
$40~\mathrm{mm}$ and diameter $L=72.5 \pm 0.1~\mathrm{cm}\simeq 10\,\sigma$. For convenience, we
define also the system radius $R=L/2\simeq 5\,\sigma$. It is convenient to define also the packing
fraction as $\phi = N(\sigma/L)^2 \simeq N/100$. We have performed experiments for three different
packing fractions $\phi= ~0.25,~0.45$ and $0.55$, which correspond here to $N=25, 45$ and $55$
particles, respectively.

An air flow is produced by means of a high precision air fan (SODECA HCT-71-6T model), which can
move air volumes of up to $1.51\times 10^4~\mathrm{m^3 /h}$. This fan is disposed so that initially
releases the air horizontally to a horizontal channel. This flow is afterwards redirected to the
vertical by means of a second channel, so that the air current impinges the particles from below. An
intermediate foam is placed within the vertical channel, with the aim of improving the upflow
homogeneity \cite{Ojha2004}. Figure ~\ref{sketch}(b) shows a sketch of the experimental set-up.

Fan power is adjusted so that all particles horizontally levitate just over the grid, thus avoiding
friction. The grid has been carefully levelled horizontally so as to avoid gravity effects. The
upflow intensity working range is delimited here by the minimum fan power necessary for the
particles to levitate and the maximum value before they start to undergo vertical displacements
(thus loosing stable horizontal alignment).  In addition, the upflow past the blades produces a
continuous particle rotation. Therefore, in our experiments, particle movement (rotation and
translation) is limited to the grid plane at all times, and is essentially two-dimensional. We
checked that deviations of the disk surface from the horizontal are less than 1\% during
experiments. Moreover, according to our measurements from a digital anemometer, the air current
intensity at the level of the grid is constant in time and is spatially homogeneous, with deviations
from the spatially averaged value of less $5\%$.

Turbulent vortexes are generated out of the von Karmann streets that are produced after a fluid flow
past an obstacle \cite{VanDyke1982,Taneda1978} (in this case, the disks). Due to the present
geometrical configuration, the typical length scale of these vortexes in our system is of the order
of the particle size \cite{Ojha2004,Lopez-Castano2021}. This effectively induces Brownian-like
movement in the particles, which in this way acquire a fluctuating component in their translational
velocities. Thus, a collisional regime is produced in the system, that so becomes \textit{fluidized}
\cite{K79}. Moreover, collisions also induce a fluctuating component in particle spins, since there
is angular momentum transfer due to friction effects upon collision \cite{Nowak1998}.

We performed a sufficiently complete set of experiments at different densities and fan power (which
results in different degrees of fluidization). Experiments were recorded for $27~\mathrm{s}\gg\tau$
($\tau$ being the typical time between collisions), for this set of experiments implies the system
is aged on average up to $10^3$ collisions per particle, in order of magnitude; i.e., steady state
conditions are achieved in all experiments during long time intervals (systems of macroscopic
particles typically achieve steady state after less than 10 collisions per particle
\cite{Montanero2000}).

For experiment recording, we used high-speed camera (Phantom VEO-410L) at a resolution of
$1280\times800$ pixels and a frame rate of 900 fps. At our working image resolution and camera
position, the pixel width is equivalent to $1~\mathrm{px}=0.9295~\mathrm{mm}\simeq 10^{-3}\sigma$,
which means that we can potentially measure particle position to a high degree of accuracy. See the
Appendix \ref{subsec:errors} for more information on the error in the experimental measurements.

Translational velocity and angular velocity (or spin) are denoted as $\mathbf{v}$ and
$\mathbf{w}=w_z\mathbf{\hat e}_z$ respectively. The unit vector $\mathbf{\hat e}_z$ is perpendicular
to the metallic grid and pointing upwards. We denote the time and spatial averages of particle
velocity and spin as $\langle \mathbf{v}\rangle, \langle w\rangle$, respectively.  We also define
the ensemble averages

\begin{align}
  \label{eq:kinetic_energies}
  & T_t(r) = (m/2)\langle(\mathbf{v} - \langle\mathbf{v}\rangle)^2 \rangle, \\
  & {T_r^*}(r)=(I/2)\langle (\mathbf{w}-\langle\mathbf{w}\rangle)^2 \rangle,
    \quad {T_r}(r) = (I/2)\langle w^2 \rangle,
\end{align}
which denote fluctuating translational kinetic energy ($T_t(r)$), fluctuating rotational kinetic
energy ($T^*_r(r)$) and rotational kinetic energy ($T_r(r)$).  Due to the specific geometry in the
system, the steady states only depend on the spatial coordinate $r$ (distance to system center). We
also define the corresponding spatially averaged (over the complete system) magnitudes:
$\overline{T_t}, \overline{T_r}, \overline{T_r^*}$. $\overline{T_r}$ can be considered as the
parameter that quantifies rotational activity; i.e., systematic rotation, see definition of $T_r$ in
\eqref{eq:kinetic_energies}.

\subsection{Particle position and angle tracking}
	 
\label{sec:algorithm}

Since our aim is to measure spin and velocity correlations, we first need to track particle
positions and angular displacements. For this purpose, we specifically wrote for this work a
particle position/angle tracking algorithm. For the translational part, our algorithm adapts several
functions from the {TrackPy} \cite{Allan2019} and the {OpenCV} libraries
\cite{opencv_library}. (Trackpy is an adaptation to Python language of the particle-tracking
algorithm originally developed by Crocker and Grier \cite{Crocker1996})

\begin{figure}
  \includegraphics[width=0.85\columnwidth]{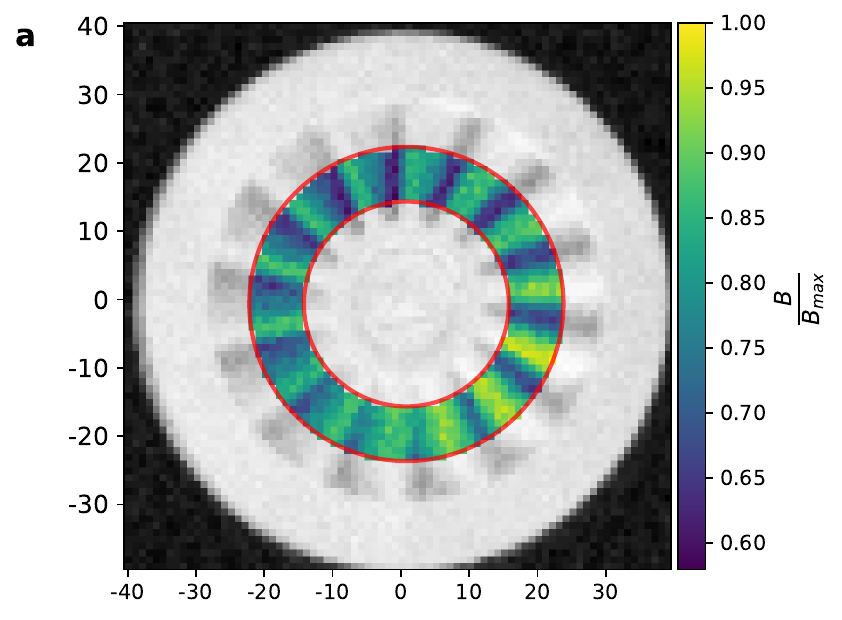}
  \includegraphics[width=0.85\columnwidth]{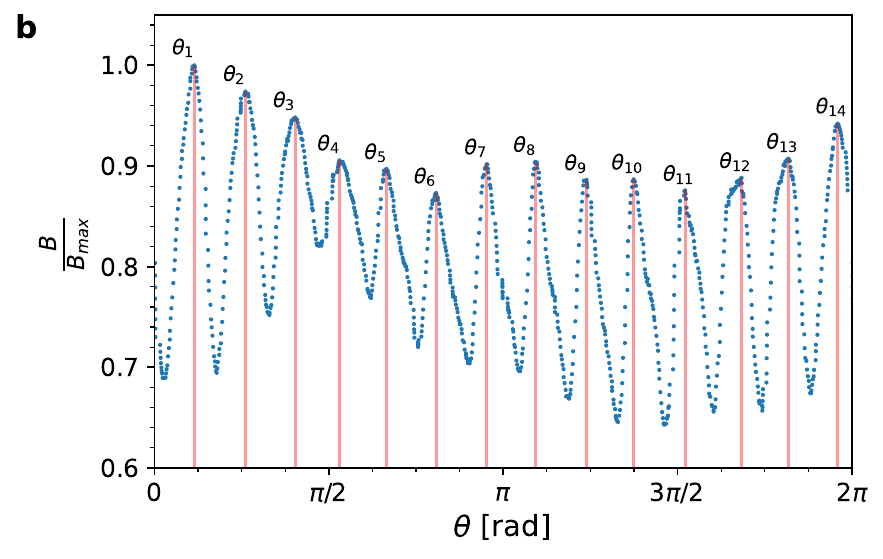}
  \includegraphics[width=0.85\columnwidth]{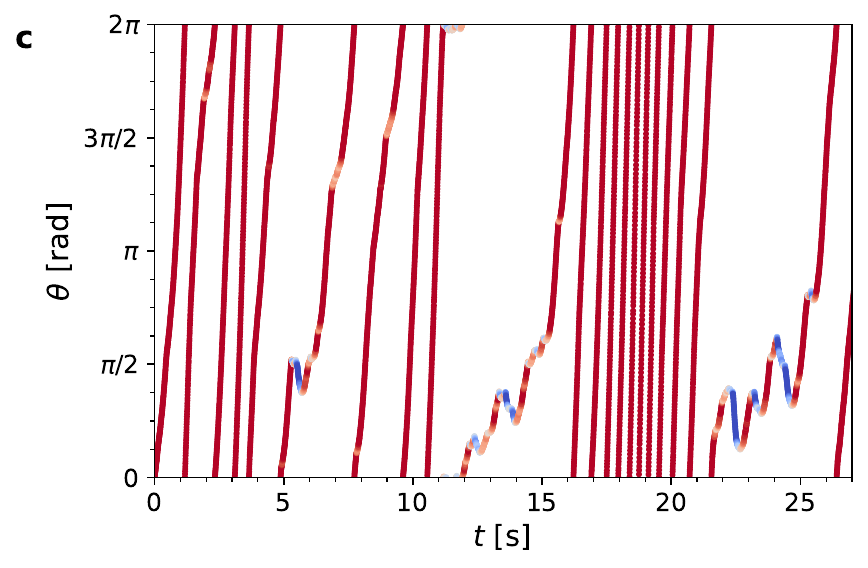}
  \caption{ (a) Image of a single disk, axis units are in pixels. The hightligthed area is the one
    used to calculate rotational spin velocity (see methods). (b) Distribution of normalized pixel
    values (y-axis), vs. angle in a frame. (c) Particle spin vs. time for a representative particle,
    in an experiment with $N=55$ particles
    and 
    $\overline{T_t}=0.91~m_p\sigma^2/\mathrm{s}^{-2}$.}
\label{fig:processing} 
\end{figure}



With respect to angle tracking, and in order to avoid periodicity effects, the camera speed is
adjusted so that the typical angular displacement between consecutive frames is not larger than half
the angle between adjacent blades (i.e., $\theta(t+dt)-\theta(t)<\pi/14$). In our experimental
configuration, this leads us to a minimum working frame rate of approximately $250~\mathrm{fps}$
(fps stands for frames per second).  Therefore, for tracking angular displacements, we have recorded
all experiments movies at $900$ fps (thus, we do not miss the large angular displacement tails of
the corresponding distribution function). This procedure will allow us, as we will see, to achieve a
high accuracy and capture finer scale details of the rotational dynamics.

The angle tracking algorithm works in the following steps:


\begin{enumerate}[label = \underline{\textbf{\textit{Step}} \arabic*:}] 
\item In each frame, at a given time $t$, and for each particle, we obtain the brightness profile
  $B_i(\theta,t)$ out of 2D spatial average of pixel value over a centered annulus (where
  $i=1\dots N$ is the particle number and $\theta$ is the polar angle with respect to an arbitrary
  reference axis).  Within this annulus, each $B_i(\theta,t)$ value is obtained from averaging pixel
  value, for $\theta$-constant pixels, from the inner to outer radii values of the annulus. In
  addition, each $B_i(\theta, t)$ is smoothed out by means of a Gaussian filter, for optimal
  accuracy. Examples of the centered annulus for brightness averaging and the obtained brightness
  profile $B_i(\theta,t)$ can be found in~\ref{fig:processing}(a) and (b) respectively.
\item After this, and for each $B_i(\theta,t)$, we determine the set of $N_b$ maxima
  $\{\theta_j(t)\}$ ($j=1\dots N_b$ is the blade index). There are always $N_b$ maxima for each
  $B_i(\theta,t)$ profile since each maximum corresponds to the tallest part of a blade (because is
  closest to a homogeneous light source, from above). Thus, each maximum is identified as the
  angular position of one of the blades, for a given frame. This is illustrated in
  Fig.~\ref{fig:processing}(b).
\item Finally, the profiles of consecutive frames are iteratively cross-correlated \cite{Press2007}
  for each particle, $B_i(\theta,t),\, B_i(\theta, t+dt)$. Specifically, cross-correlation is
  performed by using a fast Fourier transform method \cite{scipy-correlate}. From the cross
  correlation, each maximum in the set is physically identified in the next set of maxima
  $\{\theta_j(t+dt)\}$. In this way, we obtained linked $\theta_{j_1}(t)\to\theta_{j_2}(t+dt)$ and,
  therefore, the blades angular positions can be tracked throughout the entire movie. By applying
  differences between consecutive linked blade positions, we obtain instantaneous angular
  velocities. 
\end{enumerate}

The angular trajectory of a blade is represented in Fig.~\ref{fig:processing}(c). First, notice
that, since angle is periodic (between 0 and $2\pi~\mathrm{rad}$), tracks disappear and reappear in
top and bottom limits of the figure panel. Also, here symbol colors indicate counter-clockwise
(blue) and clockwise (red) particle spin (in this case, corresponds entirely to downwards or upwards
slope, respectively). Point opacity is proportional to angular velocity $w$. A completely
transparent point denotes for $w=0$; i.e., particle spin inversion. As we can see, most of the time
particle spin is clockwise, since according to the design of the blades, they rotate clockwise under
upflow. However, from time to time, spin reverses, which can be attributed to frequent particle
collisions. This is further illustrated in the Appendix, in Fig.~\ref{fig:movie} (multimedia view).


From our discussion in the Appendix~\ref{subsec:errors}, the maximum typical error from our
experimental/particle-tracking methods are $\delta r\simeq 1.7\times 10^{-3}~\sigma$ for particle
position and $\delta\theta\simeq 3.5\times10^{-3}~\mathrm{rad}$ for particle angle. This is
equivalent to about $0.01\%$ of particle diameter and $0.78\%$ of the angle between consecutive
blades, which means that errors are here very narrow and our translational and angular velocity
measurements very accurate.

\section{Results and Discussion} 
\label{sec:results}

We performed a series of measurements, varying fan power at constant packing fraction. This process
was repeated for a set of different values of packing fractions $\phi= ~0.25,~0.45$ and
$0.55$. Increasing air upflow intensity leads to higher average kinetic energy of the particles
($\overline{T_t}$) \cite{LMMRV22}. In this way, we thoroughly analyzed the behavior of translational
and rotational autocorrelations in the system, in a wide region of the relevant parameter space.

\begin{figure}[t!]
  \includegraphics[width=0.85\columnwidth]{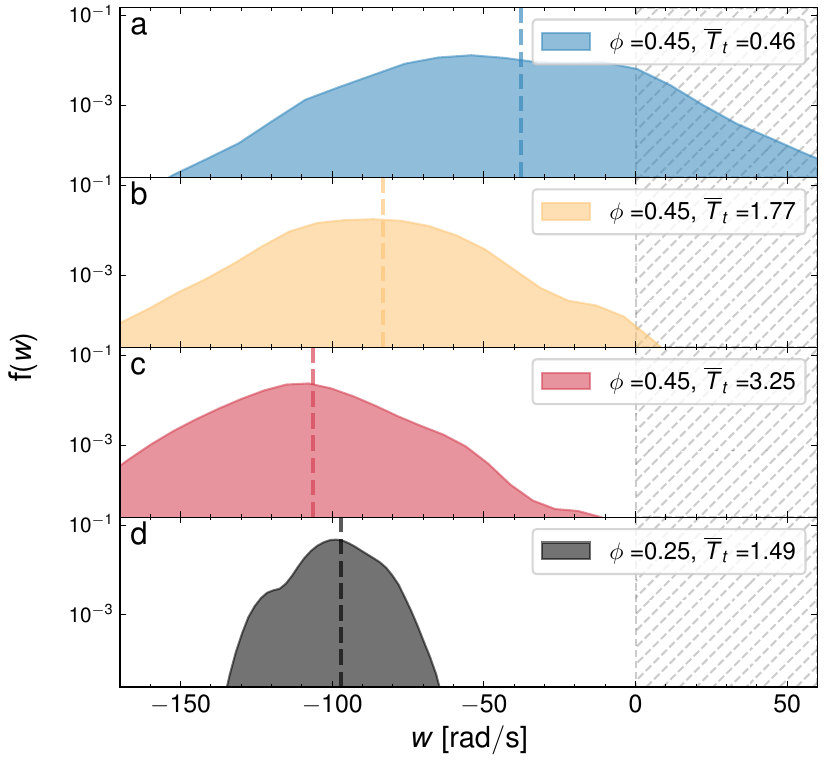}
  \caption{(a)-(c) Distribution function of particle angular velocities, $f(w)$, for 3 experiments
    with increasingly high $\overline{T_t}$, for constant packing fraction $\phi=0.45$. (d) $f(w)$
    for $\phi=0.25$ (less dense than the previous cases),
    $\overline{T_t}=1.49~\sigma^2 s^{-2}\mathrm{m_p}^{-1}$. For better visualization of the drift of
    $f(w)$ with $\overline{T_t}$ increasing, a common X-axis has been set for all four panels.}
\label{fig:distribution}
\end{figure}

Since all movie experiments were recorded under steady state conditions, we can average over all
instantaneous states, thus increasing the statistical accuracy of our analysis (we have nearly 25000
steady state snapshots).


We show in Fig.~\ref{fig:distribution} the distribution function of particles angular velocities (or
spin, $w$), $f(w)$. Particles tend to rotate clockwise by default ($w<0$), due to the particular
tilt angle of their blades, and for this reason $w$ is predominantly negative. However, as it can be
seen in certain cases, there is also a significant part of the distribution function under positive
values of $w$ due to particle collisions. In any case, as we can see in Fig.~\ref{fig:distribution},
the vertical lines signal an average spin $\langle w\rangle$ that is always negative. Moreover, the
region $f(w>0)\neq0$ decreases as fan power is increased. This points out, in our opinion, the fact
that collisions have a stronger influence in the dynamics at low $\overline{T_t}$ (i.e., at low fan
power).  Notice also that, as expected, when increasing higher upflow intensity (fan power), the
distribution shifts towards faster ranges of (negative) particle spin and at the same time its width
shrinks. As a result of both tendencies combined, absolute value of the average particle spin
increases and, more interestingly, the fraction of particles with positive spin values significantly
shrinks at intermediate fan power, finally disappearing at the largest values of $\overline{T_t}$
(or equivalenty, as we said, fan power). $\overline{T_t}$ is expressed in energy units:
$m_p\sigma^2/\mathrm{s}^{-2}$). The existence of non-vanishing $f(w>0)$ would be related thus to
emergent velocity correlations upon collision, which in this case would be mediated by frictional
effects.



\begin{figure}
  \includegraphics[width=0.85\columnwidth]{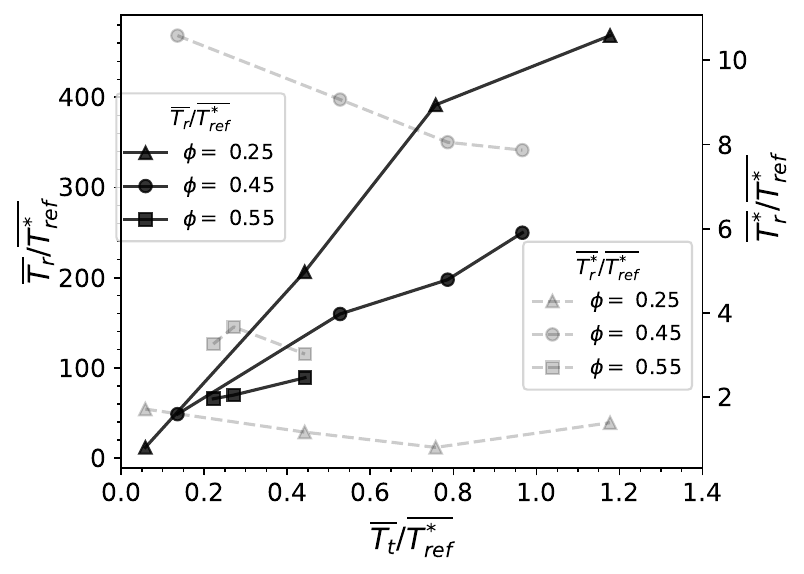}
  \caption{$\overline{T_r}$ (left Y-axis, black solid lines) and $\overline{T^{*}_r}$ (right Y-axis,
    grey dashed lines) vs. $\overline{T_t}$ for three representative values of packing fraction
    $\phi$ (legend). We use the value of $\overline{T_r^*}$ for the most dilute configuration we
    used for experiments (at packing fraction $\phi=0.03$), $\overline{T_\mathrm{ref}^*}$, as a
    reference value for dimensionalization of Y-axes.}
\label{fig:Tr_vs_Tt}
\end{figure}

In Fig.~\ref{fig:Tr_vs_Tt}, we represent $\overline{T_r}, \overline{T_r^*}$, reduced with the value $\overline{T_r^*}(\phi=0.03)=\overline{T_\mathrm{ref}^*}=3.37~m_p\sigma^2/s^2$. As we can see,
$\overline{T_r}$ increases for increasing $\overline{T_t}$. Thus, henceforth, recall that increasing
$\overline{T_t}$ implies increasing $\overline{T_r}$ and vice versa. This is expected since the
torque exerted on the disk blades, $\tau_0$, would logically be stronger for higher air current
intensity. However, it can also be seen that the rate of growth for $\overline{T_r}$ also depends on
particle density, since it tends to lower for higher particle density. This is consistent with the
fact that inelastic cooling rate due to collisions increases with density \cite{VSK14}. However, and
very interestingly, the fluctuating rotational energy $\overline{T_r^*}$ behaves exactly in the
opposite way. This clearly indicates that stronger torque $\tau_0$ (more intense particle activity)
correlates more effectively particle spin and they tend more efficiently towards a fixed (and
higher) value. This result is important since, as we will see, advances the existence of two
differentiated regimes in the dynamics: a first regime at low fluidization where collisions
decorrelate particles spins (high $\overline{T_r^*}$), and a second regime at high fluidization for
which particle spins are highly correlated and particle activity predominates.


\begin{figure}[ht]
  \includegraphics[width=0.7\columnwidth]{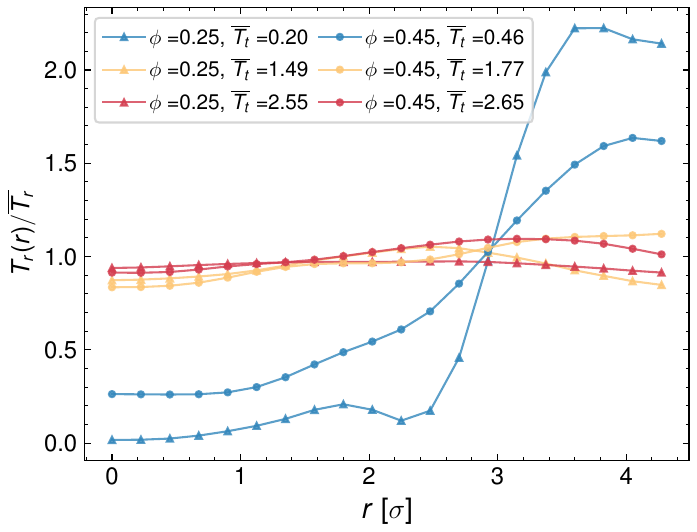}
  \caption{Radial profile of $T_r(r)/\overline{T_r}$ for two different packing fractions
    $\phi=0.25;~0.45$, (symbols ($\blacktriangle$) and ($\bullet$) respectively). For each packing
    fraction, three samples at different $\overline{T_t}$ are shown (blue, yellow and red symbols,
    respectively): $\overline{T_t}=0.20; 1.49; 2.55$ for $\phi=0.25$ and
    $\overline{T_t}=0.46; 1.77; 2.65$ for $\phi=0.45$. A step gradient is seen in the lowest
    thermalization cases, as further explained in the text.}
\label{fig:Tr_ratio}
\end{figure}


In order to further investigate on the behavior of spin fluctuations and rotational kinetic energy,
we represent in Fig.~\ref{fig:Tr_ratio} the profiles of $T_r(r)/\overline{T_r}$. This is presented
for a set of experiments at different constant values of $\phi, \overline{T_t}$. Blue curves stand
for low $\overline{T_t}$ experiments whereas orange and red denote intermediate and high
$\overline{T_t}$ respectively. Accordingly, experiments with $\phi=0.25$ are signaled with triangle
symbols and experiments series with $\phi=0.45$ are denoted with circles. The results clearly show a
transition from step function behavior at low $\overline{T_t}$ (blue curves) to a nearly flat
behavior for highly fluidized states (higher $\overline{T_t}$) (orange/red curves). Furthermore,
this transition appears consistently at low ($\phi=0.25$) and high densities (here we show
$\phi=0.45$). This step-like behavior at low fluidization indicates that non-homogeneity of the
average rotational energy is very large, with the spins in the inner part of the system rotating
significantly slower (i.e., $T_r(r<R/2)/\overline{T_r}\to 0$), but with a steep increase at
$r\simeq R/2$. On the contrary, at high and moderate fluidization, spin variations remain moderately
small and the profiles are nearly homogeneous throughout the system
($T_r(r)/\overline{T_r}\simeq 1$). This enhances the idea that there are two clearly differentiated
regimes with respect to particle spin correlations. The step-shaped form in $T_r/\overline{T_r}$ can
be attributed to inelastic cooling \cite{GoldhirschI1993}, which in the geometry of our system
induces a higher concentration of particles in the center, thereby causing more energy dissipation
at small $r$ and thus comparatively lower $T_r(r)$. The fact that this mechanism is less efficient
at high $\overline{T_t}$ is arguably due to the strengthened underlying chiral flow pattern at
higher $\overline{T_t}$, which can aid the heat flux inwards \cite{LMMRV22}.


\begin{figure*}[ht]
  \centering \includegraphics[width=0.97\columnwidth]{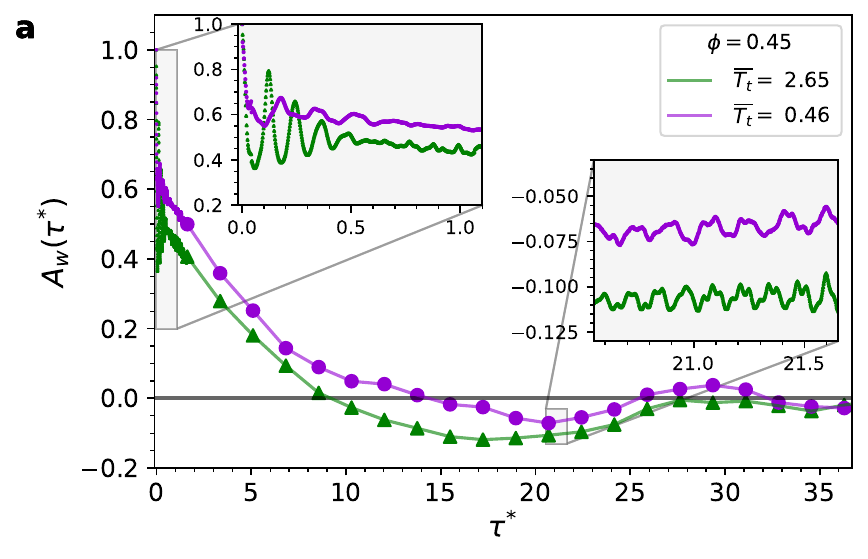}
  \includegraphics[width=0.96\columnwidth]{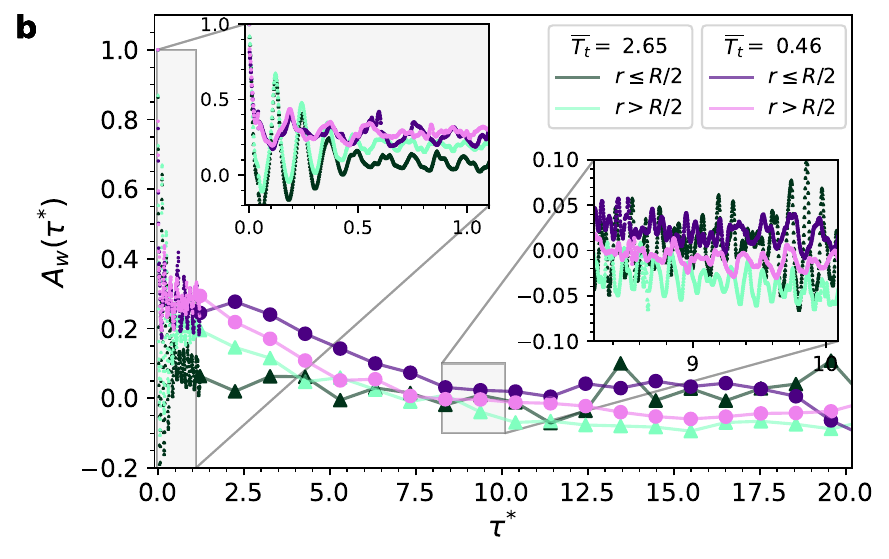}
  \caption{(a) Spin autocorrelation function, $A_{\bm{\mathbf{W}}}(\tau^*)$. We compare two cases with
    the same density ($\phi=0.45$) but different translational kinetic energies ($\overline{T_t}$ in
    units of $m_p\sigma^2 \mathrm{s}^{-2}$). Notice the small scale oscillations in left and right
    insets (at short and long delay times respectively). Their characteristic frequency is
    noticeably higher for higher rotational activity (equivalent to higher $\overline{T_t}$) (b) The
    same data sets as in the previous panels, split into two regions: $r<R/2$ and $r>R/2$. Again,
    insets zoom in for better visualization of small scale oscillations.}
\label{fig:spin_aut}
\end{figure*}

\begin{figure}[h]
  \centering \includegraphics[width=0.85\columnwidth]{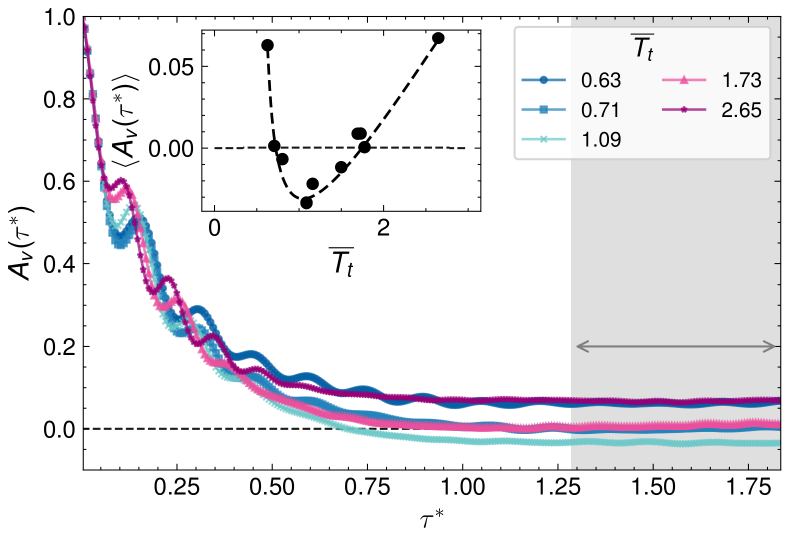}
  \caption{Velocity autocorrelation function $A_{\mathbf{v}}(\tau^*)$ for a series of experiments with
    constant packing fraction ($\phi=0.45$) and increasing $\overline{T_t}$ (in units of
    $m_p\sigma^2 \mathrm{s}^{-2}$). In the main plot we have represented the short time behavior,
    which fastly delivers a stationary plateau in all cases (signaled by the gray
    background). Notice the small scale oscillations, that have higher frequency for higher
    $\overline{T_t}$, and are rather similar to the ones in the $A_\mathbf{W}(\tau^*)$ function,
    appear here as well. The inset represents the averaged values of the velocity autocorrelation
    function $A_{\mathbf{v}}(\tau^*)$ (the average is performed over the stationary interval that is
    shaded in gray in the main panel). We added a dashed curve here, as a guide to the eye.}
\label{fig:velaut}
\end{figure}

For this reason, and with the aim of providing more detail on the evolution of the autocorrelations
\cite{Allen2017} of the fluctuating spins
$\mathbf{W}(t) = \mathbf{w}(t) - \langle\mathbf{w}\rangle$, we define the spin autocorrelation
function as:

\begin{ceqn}\begin{equation}
    A_\mathbf{W}(\tau) = \frac{\langle\mathbf{W}(t) \cdot\mathbf{W}(t+\tau)\rangle}{\langle\mathbf{W}(t)\cdot\mathbf{W}(t)\rangle},
\label{eq:autocorr}
\end{equation}\end{ceqn}
where again, $\langle\dots\rangle$ stands for averaging over all particles and steady states at
arbitrary times $t$ and fixed.


Figure~\ref{fig:spin_aut} shows the results we obtained for $A_\mathbf{W}(\tau^*)$, for an
experiment with $\phi=0.45$ and two different values of $\overline{T_t}$. Here,
$\tau^*\equiv\tau/\tau_\mathrm{ref}$, with
$\tau_\mathrm{ref}={(\overline{T_\mathrm{ref}^*}m_p^{-1}\sigma^{-2})}^{-1/2}$ is a reference time
unit we used for in Figures~\ref{fig:spin_aut},~\ref{fig:velaut}. We picked this specific value as
time unit since it refers to the most dilute system in our set of experiments (we recall,
$\overline{T_\mathrm{ref}^*}\equiv\overline{T_r^*}(\phi=0.03)$). In this way, we can refer results
to a system where particles collisions are the least significant and then refer the increase of
their influence for the rest of configurations. In the left panel, Fig.~\ref{fig:spin_aut}(a), we
represent the global space average of $A_\mathbf{W}(\tau^*)$ whereas Fig.~\ref{fig:spin_aut}(b)
represents spatial averages inside the inner and outer annular halves of the system, so as to detect
eventual differences in the correlations due to, for instance, boundary conditions. From results in
Fig.~\ref{fig:spin_aut}(a), it is clear that our new measurement method captures in great detail the
characteristic behavior of $A_\mathbf{W}(\tau^*)$ \cite{B72,Lowe1995,LMS02} since we are able to
detect short scale oscillations (which are shown in great detail in the inset).

It is very noticeable that the frequency of these tiny oscillations increases with increasing
$\overline{T_r}, \overline{T_t}$. Furthermore, these oscillations begin at very short times (of the
order of the inverse average particle spin) and that they persist at longer times (although with a
much smaller amplitude); i.e., they are present at all evolution times of the particle
dynamics. Therefore, according to all of these observations, we can outline a plausible explanation
by recalling also that particles undergo a rotational activity due to the systematic torque $\tau_0$
that the upflow yields. We think this would be the origin of these small oscillations (this in fact
being also consistent with their persistence). In effect, each time a particle undergoes a spin
fluctuation (and thus a slight decrease of $A_\mathbf{W}(\tau^*)$), then it rapidly reacts by
setting back to the value of average spin that $\tau_0$ will impose (slight increase of
$A_\mathbf{W}(\tau^*)$), and hence the oscillations. Since this torque increases for higher
$\overline{T_t}$, then it is to be expected that the tendency of the particle to setting back the
spin precollisional value (by default, closer to the average spin) is stronger as well as
faster. Hence, the increase in the frequency of the oscillations of $A_\mathbf{W}(\tau^*)$ for
increasing $\overline{T_t}$. Furthermore, we have observed that the typical frequency of
oscillations of spin autocorrelations is of the order of $10^{-2}~\mathrm{s}^{-1}$, which is
analogous to the order of magnitude of the typical values of average particle spin \cite{LMMRV22}
(we also recall again that, as Fig. \ref{fig:Tr_vs_Tt} reveals, in our setup, $\overline{T_r}$ and
hence average particle spin inherently increases with $\overline{T_t}$).

As we said, Fig.~\ref{fig:spin_aut}(b) represents a spatial analysis of the behavior of spin
autocorrelations. In particular, we represent the same experiments at strong and weak fluidization
that were shown in Fig.~\ref{fig:spin_aut}(a), but each split into two sets, each representing
$A_\mathbf{W}(\tau^*)$ averaged over only the outer ($r>R/2$) and only inner half ($r<R/2$) of the
system. Experimental data shows here that spin autocorrelations are slightly more prominent in the
outer part of the system at short times. At intermediate times, there is a crossover which renders
autocorrelations slightly stronger in the inner part of the system. This crossover seems to be also
more noticeable at higher fluidization (dark/light green data sets). However, differences between
inner and outer halves of the system are not significant. In general, notice also that spins remain
importantly anti-correlated at long times in the case of high $\overline{T_t}$ whereas for low
$\overline{T_t}$ autocorrelations remain close to zero; i.e., spins tend to decorrelate at longer
times only at low fluidization, which is reasonable because of the lower systematic torque $\tau_0$
on the particles at low $\overline{T_t}$. Significantly, smaller scale oscillations due to
systematic torque still survive at all times in both inner and outer regions, even if
$A_\mathbf{W}(\tau^*)\to 0$ at longer times (since systematic torque $\tau_0$ is always present).


	
Finally, we have also studied the behavior of the conventional velocity autocorrelation function 

\begin{ceqn}\begin{equation}
    A_{\mathbf{v}}(\tau) = \frac{\langle\mathbf{v}(t) \cdot\mathbf{v}(t+\tau)\rangle}{\langle\mathbf{v}(t)\cdot\mathbf{v}(t)\rangle}
\label{eq:autocorr2}
\end{equation}\end{ceqn}


Figure~\ref{fig:velaut} shows the results for a series of experiments with increasing
$\overline{T_t}$, at constant packing fraction $\phi=0.45$. As we can see, the decay to zero of the
$A_\mathbf{v}(\tau^*)$ curves is significantly faster if compared to that of $A_\mathbf{W}(\tau^*)$
curves (Figure \ref{fig:spin_aut}). In effect, this decay is typically observed within time
intervals of less than $0.5~\mathrm{s}$, as compared to the $5~\mathrm{s}$ that spin
autocorrelations take to fall to zero. This would be due to the fact that particle collisions are
more effective in decorrelating translational velocities since spins tend to remain correlated upon
collision due to the presence of rotational activity (i.e., the systematic torque $\tau_0$).

More importantly, we have observed also small scale oscillations in the $A_\mathbf{v}(\tau^*)$
function. Here again, the oscillation frequency is increasing for increasing $\overline{T_t}$ (and
hence, for increasing rotational activity, $\overline{T_r}$). This is, in our opinion, a
particularly interesting result since it implies that: a) linear and angular momentum exchange upon
collision would transmit these correlations oscillations to the translational degrees of freedom,
thus putting in evidence the interplay between translations and rotations in chiral particles; and
b) chirality enters, at least to some degree, into the translational degrees of freedom via velocity
correlations. The latter result is in fact in agreement with previous observations of velocity-spin
correlations, in the context of chiral flow transitions in this type of fluid \cite{LMMRV22}. With
respect to the former, this leads to the well known evidence that chiral particles can develop at
the same time chiral flows. However, we noticed that oscillations of the velocity autocorrelations
are also present when there is no persistent chiral flow pattern (i.e. even if particles do not
persistently move along circular trajectories \cite{LMMRV22}), which discards particle orbitation
along flow lines as the origin of this oscillatory behavior. It is worth to point out also that the
short-time oscillations of $A_\mathbf{v}(\tau^*)$ have been observed also in other chiral systems
very recently \cite{Zhang2020,DLJ23}, which indicates that this effect in oscillation mechanism
should be rather generic for chiral particles. Precisely because of this, it is reasonable to expect
the oscillations of the angular autocorrelations ($A_\mathbf{W}(\tau^*)$) to be generic for chiral
matter as well. All of which is now being first reported here.

Additionally, we must also highlight that the two experiments with highest and lowest
$\overline{T_t}$ display a slower and less effective decay of the autocorrelation towards zero, with
$A_\mathbf{v}(\tau^*)$ hovering around $0.1$ for a long time interval. To explain this behaviour we
must make reference to a previous work regarding the chiral fluid flow in this kind of system
\cite{LMMRV22}. The two extreme cases in Fig.~\ref{fig:velaut} correspond to strong chiral flow
vortexes, thus inducing an increase in $A_\mathbf{v}(\tau^*)$. In contrast, the intermediate case
$\overline{T_t} = 1.16$ shows negative velocity autocorrelation values for a long time which
indicates strong decorrelating effects (which corresponds to a complex phase in which multiple small
vortices of both chirality signs develop with persistent positions in time \cite{LMMRV22}). The
$A_\mathbf{v}(\tau^*\to\infty)$ values are plotted against $\overline{T_t}$ in the inset of
Fig.~\ref{fig:velaut}. As we can see, $A_\mathbf{v}(\tau^*\to\infty)$ vs. $\overline{T_t}$ shows a
pseudo parabolic behavior, with the highest values corresponding to the extreme cases of
$\overline{T_t}$ and with a negative value (anti-correlation) of the absolute minimum, at
$\overline{T_t}\simeq1.09$.

\section{Conclusions}
\label{sec:conclusions}

	
We have described in detail the statistical properties of angular velocities in a two-dimensional
fluid of disks with rotational activity. As we have shown, the dynamics of a this type of fluid has
very distinctive features due to its rotational activity \cite{Han2021}. The particle spin
distribution function, the profile of the rotational kinetic energy, the spin autocorrelations and
their radial profiles, all show the existence of two clearly different dynamical regimes. At low
fluidization (which in our system implies low rotational activity), spins are highly de-correlated,
except at short times; and rotational energy is strongly nonuniform. At strong fluidization (hight
rotational activity), by contrast, the average rotational kinetic energy remains always nearly flat,
with the spins becoming anti-correlated at long times. These two regimes consistently occur, for all
densities, at strong and weak particle activity. In summary, our results highlight the double nature
of the spin dynamics in a two-dimensional fluid of active rotors.

Rotational activity, which here can be quantified by means of the average rotational kinetic energy
$\overline{T_r}$, emerges as the fundamental control parameter for the observed phenomenology. In
fact, both $\overline{T_r}$ and average translational kinetic energy $\overline{T_t}$ increase
monotonically as upflow intensity is increased, see Figure~\ref{fig:Tr_vs_Tt}. For this reason,
$\overline{T_t}$ can be viewed in this case also as the control parameter. This is in fact
consistent with previous observations that this is the governing magnitude in the diffusive and
chiral flow behavior in this type of active fluids \cite{LMMRV22,VLR22}. 
 
The accuracy of our experimental methods, developed here for disk angle tracking, allow us to unveil
in great detail the existence of small scale oscillations in the autocorrelations of active rotors
($A_\mathbf{W}(\tau)$). These oscillations, as we commented, should appear as a consequence of the
permanent tendency of particle spin to get back to a particular value. This value is set by the
systematic torque $\tau_0$ that causes the active rotation in the chiral disks. In our case, the
driving mechanism behind this torque $\tau_0$ is the air upflow, but any other mechanism leading to
analogous effects of systematic rotation should in principle lead to the same structure of angular
velocity correlations. Since these oscillations have already been observed in analogous but
different systems of chiral fluids, we think it is reasonable to expect this oscillatory feature to
appear generically in chiral matter.

As we already observed, analogous oscillations of $A_\mathbf{v}(\tau)$ have been measured as well
very recently in chiral glasses \cite{DLJ23} and chiral fluids \cite{Zhang2020}, with also a
tendency to display higher frequency for increasing rotational activity, like in our significantly
more sparse chiral fluid. Additionally, oscillations in the time behavior of force autocorrelations
in particulate systems with odd diffusion (such as a fluid of chiral particles \cite{VLR22}) has been
recently observed \cite{KVSMS23}. However, the origin of these oscillations in autocorrelations has
not been yet thoroughly disclosed. Furthermore, to our knowledge, oscillations in the angular velocity
autocorrelation function had not been reported, in the context of chiral fluids. 

On the other hand, and according to our previous work, translational velocities of chiral particles
are statistically correlated to their rotational dynamics, with these mixed correlations having a
strong impact in the chiral fluid dynamics \cite{LMMRV22}. Together with this, we need to recall the fact that autocorrelations
have an a characteristic frequency in the range of particle spin (as we reported here in
Figure~\ref{fig:spin_aut}). Thus, interestingly, we find now evidence
that the actual origin of the oscillations  velocity autocorrelations, $A_\mathbf{v}(\tau)$, may
be found in the oscillations of the angular part of the autocorrelations, $A_\mathbf{W}(\tau)$. This
result can encourage new research, helping to cast light on the origin of the observed dynamics in a
variety of chiral fluids.


Furthermore, reports on the existence of persistent oscillations in the autocorrelation function of
angular velocities ($A_\mathbf{W}(\tau)$) are very rare, even beyond context beyond chiral
matter. According to standard hydrodynamic theory \cite{B72}, long-lived angular velocity
autocorrelations are possible due to a drag force from a surrounding fluid (in our case, it would be
the air), but no oscillations have been predicted so far. Nevertheless, oscillations for angular
autocorrelations have been reported in a single molecule (due to the existence of transversal
modes) but they are only short-lived (in comparison with the typical decorrelation time)
\cite{LMS02}. To the best of our knowledge, persistent oscillations in angular autocorrelations
have been reported previously only for a case of bonded particles (where on the other hand the
angular displacements are necessarily correlated) \cite{WKC15}, and for a case of Brownian
oscillator \cite{BPW96}. In the former case, in fact, oscillations may be due to the difference
between the image capture frequency and the intrinsic vibration frequency of the system, which is
ruled out in our experiment because rate of imaging in our experiments is at least two orders of
magnitude higher than spin rate (see Appendix \ref{subsec:algorithm_appendix} for more detail on
this subject).

Particle collisions are likely to have an effect on these correlations as well. For this reason, we
think it would be interesting for future work to analyze the behavior of a single disk rotor, in
order to isolate the influence of systematic rotation and particle collisions. Further work in this
sense is currently in progress. Another interesting direction for future work would to analyze the
properties of the angular velocity autocorrelation function for active rotors with different shapes.


\section*{Acknowledgments}
We acknowledge funding from the Government of Spain through Agencia Estatal de Investigaci\'on
(AEI), project no.  PID2020-116567GB-C22. F. V. R. is also supported by the regional Extremadura
Government through project No. GR21091, partially funded by the ERDF. A.R.-R. also acknowledges
financial support from Consejer\'ia de Transformaci\'on Econ\'omica, Industria, Conocimiento y
Universidades de la Junta de Andaluc\'ia through post-doctoral grant no. DC00316 (PAIDI 2020),
co-funded by the EU Fondo Social Europeo (FSE).

\section*{Data Availability Statement}
The data that support the findings of this study are available author upon request to the
corresponding author. The exact version of the particle tracking codes that were used for this work
are available in the following repository: \texttt{https://zenodo.org/record/7226255}.
	
\appendix*
\section{}

\subsection{Angular displacement tracking algorithms}
\label{subsec:algorithm_appendix}
An accurate measurement of particle rotation is crucial for a wide variety of applications, such as
the characterization of the tangential collisional inelasticity of macroscopic particles
\cite{VSK14,Labous1997,Jiang2020}, design of microrotor properties for their use as non-invasive
drug delivery vectors \cite{Tierno2021} or the evaluation and calibration of rotating machinery in
many industrial scenarios. Although electrical sensors are commonly used for this task, there are
scenarios in which optical techniques might be preferable, for instance: situations where the
placement of physical electrical devices is difficult, dangerous, or expensive (such as extreme
temperature/pressure environments), or for determining the rotation of very distant objects, for
example, drone-based wind turbine characterization; it can be also applicable to very small objects,
for which on board installation of physical sensors is difficult (proteins, microgears, etc.).

In previous experimental works, several methods are routinely employed to extract particle angular
velocity. A simple approach would typically employ a stroboscope \cite{Lee1996} or an on-board
mounted sensor (tachometer) \cite{Lu2021}. In camera-based experiments, another common approach is
to analyze the asymmetrical features of the object \cite{Scholz2018} or to put a tracer mark on the
particle \cite{Grasselli2015,Workamp2018}. Angular displacements are computed in this case by
measuring the mark angle relative to a fixed axis, which passes through the particle center. This
method requires the identification of two points: the particle center and mark position. In the case
of spheres, more complex methods have been described
\cite{Barros2018,Zimmermann2011,Hagemeier2015a}, often making use of twin cameras to obtain a three
dimensional perspective \cite{Tsai1987}.
	
However, depending on the precise experimental conditions, physical particle marking might not be
feasible. Thus, it is useful to develop a method which does not require marking. For instance, one
such method may consist in looking at the pixel intensity arrays of the particle for two consecutive
frames \cite{Adrian1991,Helminiak2018}. These two frames are iteratively cross-correlated
\cite{Papoulis1978}, in each step rotating the second array a certain angle $d \theta$. This
operation results in a probability distribution of scalar coefficients $P(\theta)$, indicating the
similarity between the original frame and the rotated second frame. The corresponding maximum
likelihood estimate of a Gaussian signals the true angular displacement. This method has the
drawback that computational cost quickly increases for increasing angular resolution (since
computing cross-correlation (convolution) is very time consuming) \cite{Helminiak2018}.

For this reason, we implement a modified convolutional method directly into brightness profiles
(which are 1D arrays), as described in section~\ref{sec:algorithm} instead of using complete frames
(2D arrays). Although vectorization \cite{VanderWalt2011} does not decrease the number of
operations, it allows for their computation within a common series of CPU cycles. This shrinks the
execution time, in comparison with a loop of independent cross-correlations of particle images,
which is the main alternative method \cite{Helminiak2018}. We have empirically measured the
computational complexity of our algorithm, which grows with the number of particles present in the
video as $\mathcal{O}(N)$. We have implemented this algorithm in Python (a repository is maintained
\cite{github}, where the most recent version of the code can be downloaded) running on an
Intel\textsuperscript{\tiny\textregistered} Xeon\textsuperscript{\tiny\textregistered} Gold 6240 CPU
@ 2.60GHz. In order to compare our execution time with that of previous works, we used 32 threads.
Our tests show that we can process a typical experimental realization in under 30 minutes, which is
faster than cross correlation-based methods \cite{Helminiak2018}, up to a factor of
$10^2$. 
With further parallelization and optimizations, we are confident that this angular velocity tracking
could be applied in real-time, in fact, we have checked that for $N$ between 1 and 3 disks we can
run the angular detection code in real-time. This can be very useful for a wide variety of
industrial and research applications.

\subsection{Estimation of experimental errors}
\label{subsec:errors}

We now analyze the possible error sources of angular velocity measurement for our method. In this
case, uncertainties such as motion blur and static errors \cite{Savin2005} (coming from lightning
and camera noise) \cite{Feng2011a,Sciacchitano2019} are less important when compared to the main
limitation of the present approach: the limited number of pixels used for detecting brightness
maxima.

From our analysis (which we will further expand below), we estimate that the maximum typical error from our
experimental/particle-tracking methods are $\delta r_\mathrm{dyn}\simeq 1.7\times 10^{-3}~\sigma$
for particle position and $\delta\theta_\mathrm{res}\simeq 3.5\times10^{-3}~\mathrm{rad}$ for particle
angle. Note the degree of accuracy of our measurements: angular error is equivalent in this case to
only a 0.78\% of the angle between consecutive blades and positional error is less than a 0.2\% of
particle diameter.

Subscripts indicate the origin of the error source.  In the case of positions, the source of this
maximal typical error is the so-called dynamical error. In the case of angles, the source is the
camera resolution itself (which indicates our method is \textit{as good as it can be}) for the
high-speed camera model we are using.
We discuss below on the different error sources, and the process for estimation of the figures for
maximum typical error given above.

\textbf{Static error}: Static error is usually defined as the underlying noise in the measurements, things like
light flickering or camera calibration inaccuracies. We took videos of the particles in a static
situation and measured their coordinates and angular positions in order to search for these
deviations, we have estimated from recordings of a still disk (standard deviation) the effects of static error to be
around $\delta r_\mathrm{sta}=0.06~\mathrm{px}\simeq 6\times 10^{-5}~\sigma$ for positions and
$\delta\theta_{\mathrm{sta}} \simeq 1.9\times 10^{-5}~\mathrm{rad}$ for angles, which are negligible.

\textbf{Dynamic error}: Image dynamic blur is a phenomenon cause by the fact that particle coordinates
are not recorded at a precise instant, but instead, during a finite acquisition time interval. The lower the
shutter time the lower the error. In these experiments, we used a shutter time of
$\Delta t_\mathrm{exp}=1.5\times 10^{-3}~\mathrm{s}$. We also recall that therefore, the error
\cite{Savin2005} caused by this finite acquisition time should be
$\delta\theta_{\mathrm{dyn}} =\Delta\overline\omega (\Delta t_\mathrm{exp}/\Delta t_\mathrm{fps})\simeq
7.5\times 10^{-5}~\mathrm{rad}$. We also estimated the mean translational error due to motion blur to be 
$\delta r_{\mathrm{dyn}} = \overline{\Delta r} (\Delta t_\mathrm{exp}/\Delta t_\mathrm{fps})\simeq
1.7\times 10^{-3}\sigma$, where $\Delta\overline r, \Delta\overline\omega$ are typical position and
angle displacements between consecutive frames.


\textbf{Sub-sampling error}: This source of uncertainty is often overlooked, one could well detect
particle positions and angles with perfect accuracy but, if the sampling rate is very low, the real
trajectories can not be followed. Therefore, we must make sure that the videos are recorded at a
sufficiently high framerate so that the angular trajectories of the disks recorded are as close to
the ground-truth as possible. However, there are trade-offs, we can not increase the frame
acquisition rate \textit{ad infinitum} since the amount of memory available in the camera is
limited. In order to quantify this, we define the translational sub-sampling error as:
$\delta r_\mathrm{sub}(n)\equiv(1/\mathcal{N}')(\mathcal{L}(1)-\mathcal{L}(0))$, where
$\mathcal{L}(n)=\sum_{i=1}^{\mathcal{N}'}|\mathbf{r}_i-\mathbf{r}_{i-1}|$ is the trajectory length
as if recorded at a fraction $n$ of the base frame rate of 900 fps. Here, $\mathcal{L}(0)$ is the
length of the trajectory measured from a Savitzky-Golay interpolation out the movie at 900 fps, at
order 3 and step 5 \cite{Lopez-Castano2021}. Therefore $\mathcal{L}(n)$ represents the length of the
measured trajectory for a framerate $900/n~\mathrm{fps}$ and $\mathcal{N}=n\,\mathcal{N}'$ is the
total number of measured positions in a trajectory at the base 900 fps rate. In our case, after a
careful analysis over all experiments and trajectories, we found that the sub-sampling errors for
the base frame rate are $\delta r_{\mathrm{sub}}(1) \simeq 4.6\times 10^{-4} \sigma$ for
translations and $\delta \theta_{\mathrm{sub}}(1) \simeq 3.3. \times 10^{-3} ~\mathrm{rad}$ for
rotations.
        	
\textbf{Limited camera resolution}: Due to the finite resolution of the camera, the inner and outer
radii of the annulus where brightness profiles $B_i(\theta)$ are measured (see red circumferences in
Fig.~\ref{fig:processing}(a)) are, in our case, $r_\mathrm{in} = 15~\mathrm{px}$ and
$r_\mathrm{out} = 23~\mathrm{px}$ respectively (in units of image pixel length), meaning that
the brightness annulus occupies $N_p = 955$ pixels (see Fig.~\ref{fig:processing}(a)), each
corresponding to a different value of $\theta$. As a consequence, the complete angle interval
($2\pi$) is subdivided into 955 segments (the ones defined by the set of angle points in
Fig.~\ref{fig:processing}(b)). Therefore, the angle resolution is, at best:
$\delta\theta = 2\pi/N_p$. This implies that the error of the average of the inter-frame differences
over all blades angles ($\langle\Delta\theta(t)\rangle$) is
$\delta\langle\Delta\theta\rangle_{\mathrm{res}} =
\sqrt{\frac{2}{N_b}}\frac{2\pi}{N_p}=\pm~3.5\times 10^{-3}$ rad. 
This yields an error ()relative to a blade section) of
$\frac{\delta\langle\Delta\theta\rangle_{\mathrm{res}}}{(2\pi/N_b)}=0.78\%$, a value that for our
purpose is quite satisfactory. Similarly, since $1~\mathrm{px}\simeq 10^{-3}~\sigma$, and the
particle tracking method has a limit resolution of 0.1 px, the camera resolution error on particle
position is around $\delta r_\mathrm{res}\simeq 10^{-4}~\sigma$.

\textbf{Other sources of error}: We also tried to limit other sources of error as
much as possible, for example, regarding the issue of 3D-printing reproducibility, we have measured
the mass distribution of our whole set of disks with a precision scale and found that they all
deviate less than $0.1~\mathrm{g}$ from the mean mass. We also took measures to ensure the
homogeneity of the driving air flow such as installing a porous polyurethane layer below the arena,
that way fluctuations in the upflow intensity are limited to $\pm 5\%$ of the mean value.

\begin{figure}[ht]
  \centering \includegraphics[width=0.90\columnwidth]{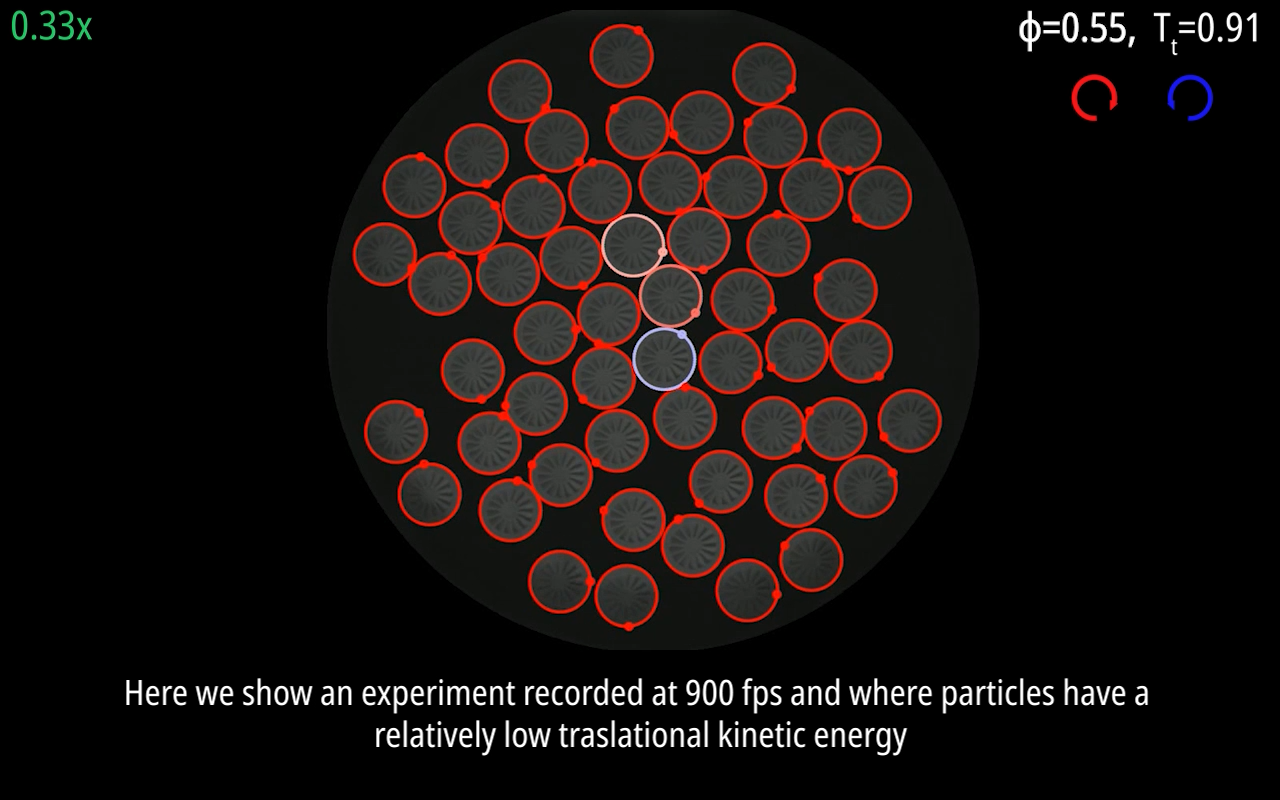}
  \caption{Movie of an experiment (multimedia view) at real-time speed. Each particle is highlighted by a surrounding
    color circle. Changes in color indicate changes in the spin sign. Blue stands for reverse spin
    (counter clockwise in this case), whereas red indicates spin in the natural direction imposed by blade
    tilt (which is the most commmon state at all times, and in this case corresponds to clockwise
    rotation). Packing fraction $\phi=0.55$ and global mean translational kinetic energy
    $\overline{T_t}=0.91~\mathrm{m}\sigma^2\mathrm{s}^{-2}$\label{fig:movie} }
\end{figure}

Finally, the reader can visually check the accuracy of spin reversal detection for our measurements
in Fig.~\ref{fig:movie} (multimedia view). In this experimental movie, a mark that tracks the
detected angle resulting from our algorithm is dynamically superimposed on each particle. It is
noticeable for instance that particle spin sign reversals (signaled in blue in
Fig.~\ref{fig:processing}c) are detected with no error.

For details on the methodology we followed for determination of the translational parts of the
errors, the reader may refer to our previous study on an analogous set-up \cite{Lopez-Castano2021}.

\section*{References}
\bibliography{spin_tn}
	
\end{document}